\documentclass[aps,prd,reprint,nofootinbib,twocolumn]{revtex4-2}

\usepackage[english]{babel}
\usepackage{amsmath}
\usepackage{amsfonts}
\usepackage{graphicx}
\usepackage[colorlinks=true, allcolors=blue]{hyperref}
\usepackage{cancel}
\usepackage[export]{adjustbox}
\hypersetup{breaklinks=true}

\usepackage{footmisc}

\newcommand{\VSI}{Van Swinderen Institute for Particle Physics and Gravity,\\ University of Groningen,
Nijenborgh 4, 9747 AG Groningen, The Netherlands}

\newcommand{\mac}{Department of Physics and Astronomy,
Macalester College,
Saint Paul MN 55105-1899, U.S.A.}

\begin{document}
\title{Hybrid Goldstone Modes from the Double Copy Bootstrap}
\author{Yang Li}
\author{Diederik Roest}
\affiliation{\VSI}
\author{Tonnis ter Veldhuis }
\affiliation{\mac}

\begin{abstract}
\noindent
We perform a systematic classification of scalar field theories whose amplitudes admit a double copy formulation and identify two building blocks at 4-point and 13 at 5-point. Using the 4-point blocks simultaneously as bootstrap seeds, this naturally leads to a single copy theory that is a gauged {nonlinear sigma model}. Moreover, its double copy includes a novel theory that can be written in terms of Lovelock invariants of an induced metric, and includes {Dirac-Born-Infeld} and the {special Galileon} in specific limits. The amplitudes of these Goldstone modes have two distinct soft behaviour regimes, corresponding to a hybrid of non-linear symmetries.
\end{abstract}

\maketitle

\section{Introduction}

\noindent
The double copy framework manifests a remarkable connection between the unique (at lowest order in derivatives) interacting theories of spin-1 and spin-2: the amplitudes of General Relativity (GR) can be written as the squares of specific color-dual kinematic numerators that define  Yang-Mills (YM) \cite{Bern:2008qj, Bern:2010ue}. This connection has its origin in open-closed string duality \cite{Kawai:1985xq} and is closely related to the scattering equations approach of  \cite{Cachazo:2013gna,Cachazo:2013hca}. The double copy has since been extended to include supersymmetric theories and loop level, as reviewed in \cite{Bern:2019prr,Bern:2022wqg}, as well as scalar field theories with enhanced soft limits that generalise the Adler zero and hence can be seen as Goldstone theories \cite{Cheung:2014dqa, Cheung:2017pzi}.

A natural question regards the uniqueness of the kinematic numerators: are higher-derivative corrections encoded in other color-dual kinematic numerators? For the color-dual kinematic numerator of YM, there is a single additional possibility at 3-point (while at 4-point, there are already 8 different tensorial structures \cite{Bern:2017tuc, Carrasco:2019yyn}) that generates the unique $F^3$ correction. Using this as a seed interaction, double copy compatibility at 4-point then implies the inclusion of a $F^4$ term. Moreover, following the same logic at 5-point requires the further quartic term $D^2 F^4$ \cite{Carrasco:2022lbm}. It was conjectured to go up to all derivatives, leading to a UV complete series that is part of the bosonic open string amplitude \cite{Carrasco:2022lbm,Elvang:2018dco}.

A related result was found very recently for higher-derivatives to a specific scalar field theory, the nonlinear sigma-model (NLSM) \cite{Brown:2023srz}. Again, higher-derivative corrections to the 4-point seed interactions were found to be constrained by higher-point consistency. The only known theory that satisfies these constraints at all order (apart from the NLSM itself) is Z-theory, again with an infinite tower of derivatives\footnote{Similar infinite series were also found in generalisations of the KLT kernel of the bi-adjoint scalar theory in \cite{Chen:2023dcx}; it would be interesting to investigate how these relate to the current results.} \cite{Carrasco:2016ldy}.

In this Letter, we perform a related analysis for scalar field theories with Goldstone modes. The crucial difference with \cite{Brown:2023srz} is that we do not restrict ourselves to 4-point {\it contact} interactions. As we will show, this allows for a unique additional {\it exchange} interaction. Similarly, we classify all possible double copy scalar seeds at $n=5$ and find 29 independent structures, of which generate 13 physical amplitudes. 

Using a linear combination of the two 4-point seed interactions, we then employ the bootstrap procedure to construct theories for Goldstone modes with a hybrid character: while the entire theory has a particular soft degree $\sigma_{\rm min}$ (where the soft degree $\sigma$ is defined as $A_n\sim {\cal O}(p^{\sigma})$ when an external momentum $p$ becomes soft), it contains a subsector that is defined by having $\sigma_{\rm max} = \sigma_{\rm min} +1$ instead. We present examples of a single and a double copy: a gauged version of the NLSM with $\sigma_{\rm min} = 0$ and a particular higher-derivative extension of Dirac-Born-Infeld (DBI) with $\sigma_{\rm min} = 2$. As the latter is formulated in terms of Lovelock invariants, we refer to it as DBI-Lovelock. 

In contrast to the results of \cite{Carrasco:2022lbm, Brown:2023srz}, we find no need for infinite sets of quartic higher-derivative corrections; in this sense, our results are more akin to the extended DBI theory \cite{Cachazo:2014xea, Low:2020ubn, Kampf:2021tbk}. We  provide an interpretation for this difference in the conclusion, and outline further implications and generalisations.

\section{BCJ Representations}

\noindent
The double copy or BCJ approach \cite{Bern:2008qj,Bern:2010ue, Bern:2019prr,Bern:2022wqg} has identified a number of field theories, famously including GR and YM, whose amplitudes can be rewritten in terms of a sum over $(2n-5)!!$ trivalent diagrams:
 \begin{align}
     A_n = \sum_{\text{trivalent}}\frac{N\tilde{N}}{D} \,.
     \label{double copy amplitude}
 \end{align}
The denonimator in the above sum consists of the propagators for every diagram, while the numerator instead is the product of two so-called BCJ numerators that encode the characteristics of the particles in the scattering process. While each trivalent diagram has an associated kinematic numerator, only $(n-2)!$ of these are independent; a convenient basis for these is provided by the Del Duca-Dixon-Maltoni (DDM) basis \cite{DelDuca:1999rs}.

An important and arguably the simplest example is given by the colour factors that consist of products of structure constants $f_{abc}$. At multiplicity $n$, these are given by a product of $n-2$ structure constants:
 \begin{align}
  N_{abc \ldots} = f_{ab}{}^{x_1} f_{x_1c}{}^{x_2} f_{x_2 \ldots} \ldots \,. \label{structure-constants}
\end{align}
When viewed as a representation of the permutation group $S_n$, the above numerators satisfies the nested commutator structure \cite{deNeeling:2022tsu}
\begin{equation}
-N_{abcd\dots}=N_{bacd\dots}=N_{c[ab]d\dots}=N_{d[[ab]c]\dots}=\ldots \,,
\label{Jacobi id}
\end{equation}
which will be referred to as generalised Jacobi identities.  Moreover, the colour factors are even or odd under reflection,
\begin{equation}
N_{abcd\dots}=(-)^nN_{\dots dcba} \,.
\label{reflection inva}
\end{equation}
We have identified which  irreducible representations (irreps) of $S_n$ the above constraints correspond to, with the dimensions of these irreps adding up to $(n-2)!$ in every case, see table~\ref{table:BCJ irrep}.

\begin{table}
\centering
\begin{tabular}{||c | |c|c||} 
 \hline
$n$ & Kinematic numerators \\ [0.5ex] 
 \hline \hline
 4 &  $[2,2]$  \\ 
 \hline
 5 &  $[3,1,1]$ \\
 \hline
 6 & $[4,2]$, $[3,1,1,1]$, $[2,2,2]$ \\
 \hline
 7  & $[5, 1, 1]$, $[4, 2, 1]$, $[3, 3, 1]$, $[3, 2, 1, 1]$, $[2, 2, 1, 1, 1]$ \\
 \hline
 8 &  \begin{tabular}{@{}c@{}} $[6,2]$, $[5,2,1]$, $[5,1,1,1]$, $[4,4]$, $[4,3,1]$, $2\times [4,2,2]$,  \\ $[4,2,1,1]$, $[4,1,1,1,1]$, $2\times[3,3,1,1]$,  \\ $[3,2,2,1]$, $[3,2,1,1,1]$, $[2,2,2,2]$, $[2,2,1,1,1,1]$\end{tabular}\\
 \hline
\end{tabular}
    \caption{\it BCJ-compatible  numerators as irreps of $S_n$.}
\label{table:BCJ irrep}
\end{table}

Instead of structure constants, we will be interested in  color-dual kinematic numerators that only contain Mandelstam variables (scalar numerators for short). These are relevant for scalar field theories: in single scalar field theories, the particles only carry momentum information and thus Mandelstam invariants. Moreover, in multi-scalar field theories such as the NLSM with multiple flavours plus higher-derivative corrections, the colour information factorises and thus one of the two BCJ numerators again only involves Mandelstam variables. 

The possibilities can be phrased in terms of $S_n$ representations. The set of $n$ external momenta forms the so-called standard irrep $[n-1,1]$ with dimension $n-1$. Lorentz invariants then consist of inner products of momenta and live in the irrep $[n-2,2]$ with dimension $\tfrac12 n (n-3)$; these correspond to the Mandelstam invariants. Moreover, we work in general dimensions and hence are not affected by Gram determinant considerations that reduce the number of independent Mandelstam variables.

The above approach reduces the classification of scalar numerators to a representation theory problem\footnote{It can also be phrased in representation theory of the cyclic group instead, with all cyclic invariants generating an overcomplete basis of scalar numerators \cite{Bonnefoy:2021qgu}.} of $S_n$: for the number of scalar numerators at a given multiplicity $n$ and at a given order $p$ in Mandelstam variables, one simply calculates the symmetric product of $p$ irreps $[n-2,2]$ and decomposes this into $S_n$ irreps. A comparison with the BCJ-required irreps of table \ref{table:BCJ irrep} then directly gives the number of possible scalar numerators at this order. 

\begin{table}
\centering
\begin{tabular}{||c | |c|c||} 
 \hline
$n$ & Gauge parameters \\ [0.5ex] 
 \hline \hline
 4 &  $[1,1,1,1]$  \\ 
 \hline
 5 &  $[2,2,1]$ \\
 \hline
 6 &  $[3, 2, 1]$, $[3, 1 ,1 ,1]$\\
 \hline
 7  &  \begin{tabular}{@{}c@{}} $[4, 3]$, $[4, 2, 1]$, $[4, 1, 1, 1]$, $[3, 2, 2]$, $[3, 2, 1, 1]$,\\ $[3, 1, 1, 1, 1]$, $[2, 2, 2, 1]$ \end{tabular}\\
 \hline
\end{tabular}
    \caption{\it BCJ-compatible gauge parameters as irreps of $S_n$. }
\label{table:BCJ gauge}
\end{table}

Not all scalar numerators will contribute to the amplitude; some solutions $N$ will give a vanishing contribution to \eqref{double copy amplitude}, independent of the choice for $\tilde N$. Interestingly, these gauge solutions can also be characterised by representation theory: all scalar numerators that can be written as the product of Mandelstam variables with a specific $S_n$ irrep drop out of the amplitude. The first example surfaces at 4-point and reads
 \begin{align}
 N_{abcd} = s_{ab} G_{abcd} \,, \label{gauge-4pt}
 \end{align}
where the convention $s_{i\dots j}=(p_i+\cdots + p_j)^2$ is adopted, and $G$ is a fully anti-symmetric tensor and hence lives in the $[1,1,1,1]$. We have listed the analogous irrep requirements at higher multiplicities in table \ref{table:BCJ gauge}. For further details, see \footnote{Further details on the construction and counting of BCJ-compatible gauge numerators and a discussion of gauged NLSMs can be found in the Supplementary Material \ref{sec:appendix}, which includes \cite{Carrasco:2022sck,Johansson:2017srf,Johansson:2018ues,Cachazo:2013iea,Cachazo:2014nsa,Cachazo:2016njl,He:2018pol}.}.

\section{BCJ seed classification}

\noindent
We will now proceed to systematically classify all scalar numerators at lower multiplicities at at 4- and 5-point\footnote{At three-point it is impossible to construct scalar numerators, as there are no Mandelstam invariants; relatedly, three-point operators with derivatives can always be reformulated as four-point operators and hence give vanishing three-point amplitudes.} using representation theory.
 
At {\bf four-point}, the required BCJ irrep is the window of $S_4$ with dimension $2$. The Mandelstam invariants in this case live in the same irrep. Therefore there is naturally a linear combination of Mandelstam invariants that satisfies the BCJ constraints. An explicit construction shows that it is given by 
 \begin{align}
    N_4^{(1)}=s_{bc} - s_{ac} \,.
    \label{4pt_win_1}
 \end{align}
Moreover, at quadratic order in Mandelstam, the symmetric product of two window irreps decomposes into $[2,2] + [4] + [1,1,1,1]$  and hence there is another scalar numerator for four-point at this order. It takes the form
 \begin{align}
      N_4^{(2)}=s_{ab} (s_{bc} - s_{ac}) \,.
     \label{4pt_win_2}
 \end{align}
The expression \eqref{4pt_win_1} corresponds to the four-scalar scattering with an exchanged gluon, and \eqref{4pt_win_2} to the four-point contact interaction of the NLSM. We will refer to the linear and quadratic solutions as exchange and contact scalar numerators, respectively.

At higher orders, there are new solutions to the generalised Jacobi. However, it follows from representation theory that these are always of the form of one of the two above building blocks, multiplied by Mandelstam expressions that are separately invariant (and hence can be used to construct additional solutions to the BCJ conditions). To see this, note that the number of invariants at every order is given by the Taylor coefficients of the Molien series\footnote{This coincides with the Hilbert series for the case of invariant polynomial rings.}{\cite{benson_1993,Boels:2013jua,derksen2015computational,Henning:2017fpj}
}
 \begin{align}
    H_4^{\rm Inv}(x)=\frac{1}{(1-x^2)(1-x^3)} \,.
    \label{4-pt Hilbert series}
 \end{align}
This amounts to the statement that all invariants can be written as arbitrary powers of two primary invariants:
 \begin{align}
   I_4^{(2)} =s_{ab} s_{bc} + s_{ac}s_{bc}+s_{ab}s_{ac} \,, \quad  I_4^{(3)} = s_{ab} s_{ad} s_{ac} \,.
   \label{4pt invariant Y and X}
 \end{align}
Moreover, the number of window irreps at every order in Mandelstam is generated by 
 \begin{align}
   H_4^{\rm BCJ}(x) = (x+x^2) H^{\rm Inv}_4(x) \,.
    \label{4pt Hilbert series BCJ}
 \end{align}
All window solutions are therefore either \eqref{4pt_win_1} or \eqref{4pt_win_2} multiplied by an invariant, as also found in \cite{Carrasco:2019yyn}.

Turning to gauge parameters, the Molien series for the relevant irrep $[1,1,1,1]$ is given by
 \begin{align}
   H_4^{\rm Gauge}(x) = x^3 H^{\rm Inv}_4(x) \,,
 \end{align}
which generates the number of gauge parameters at one order higher. Indeed, it turns out that the combination
 \begin{align}
  2N_4^{(2)} I_4^{(2)} - 3 N_4^{(1)} I_4^{(3)} \,,
  \label{4pt gauge factor}
 \end{align}
is of the form \eqref{gauge-4pt} and drops out of the amplitude \eqref{double copy amplitude} for any scalar numerator $\tilde N$. The number of physical BCJ parameters is therefore given by
 \begin{align}
     H_4^{\rm BCJ}(x) - x H_4^{\rm Gauge}(x) = \frac{x}{1-x^2} + \frac{x^2}{(1-x^2)(1-x^3)} \,,
 \end{align}
generated by the linear or quadratic seed solutions \eqref{4pt_win_1} and \eqref{4pt_win_2} multiplied by quadratic and/or cubic invariants\footnote{This general solution includes the 4-point Abelian Z-theory \cite{Carrasco:2016ldy} for a specific tuning of its coefficients, as suggested by \cite{Elvang:2018dco,CarrilloGonzalez:2019fzc,Brown:2023srz}.}.

At {\bf five-point}, the story is similar but more complicated. The 5-point Hilbert series for invariants is given by 
\begin{equation}
   H_5^{\rm Inv} (x) = (1+x^6+x^7+x^8+x^9+x^{15}) / D_5(x) \,,
   \label{5pt Hilbert series invariant}
\end{equation}
with the denominator given by
 \begin{align}
     D_5(x) = (1-x^2)(1-x^3)(1-x^4)(1-x^5)(1-x^6) \,.
 \end{align}
Each factor in the denominator corresponds to a primary invariant, and each term in the numerator to a secondary invariant; for example, $(1-x^2)$ represents the contribution from a quadratic primary invariant, whereas $x^6$ represents a sextic secondary invariant. The difference is that primary invariants can appear at any power to form new invariants, while there can only be a single secondary invariant. The latter restriction is due to relations between products of invariants, referred to as syzygies \cite{benson_1993,derksen2015computational}.

The BCJ irreps, instead, are given by $[3,1,1]$ corresponding to the ``hook'' Young tableau. The Molien series for this is\footnote{These numbers were found up to order 12 in \cite{Carrasco:2021ptp}, which also includes explicit expressions for the scalar numerators.}
 \begin{align}
   H_5^{\rm BCJ}(x) = ( & x^3 + 2 x^4 + 4 x^5 + 5 x^6 + 6 x^7 + 6 x^8 + 5 x^9 \nonumber\\
   & + 4 x^{10} + 2 x^{11} + x^{12}) / D_5(x) \,,
    \label{5pt Hilbert series BCJ}
 \end{align}
and are thus given by the numerator structures multiplied by primary invariants. 

The gauge parameters in this case are generated by the irrep $[2,2,1]$; the number of such irreps at every order is generated by 
 \begin{align}
     H_5^{\rm Gauge}= (  &x^2 + x^3 + 3 x^4 + 3 x^5 + 3 x^6 + 4 x^7 + 4 x^8 \nonumber\\
     & + 3 x^9 + 3 x^{10} + 3 x^{11} + x^{12} + x^{13} ) / D_5(x) \,.
 \end{align}
However, in this case the number of distinct resulting gauge parameters (at one order higher) is somewhat smaller and given by 
\begin{align}
    &(x^3 + x^4 + 2 x^5 + 3 x^6 + 2 x^7 + 3 x^8 + 4 x^9 + 3 x^{10} \nonumber\\ 
 & + 2 x^{11}+2 x^{12} + x^{13}) / D_5(x) \,.
  \label{5pt Hilbert series gauge}
\end{align}
This difference comes about as some BCJ parameters can be split into Mandelstam times $[2,2,1]$ in multiple ways. The resulting number of physical BCJ parameters at every order is given by the difference of \eqref{5pt Hilbert series BCJ} and \eqref{5pt Hilbert series gauge} and can be written as a sum of fractions with positive coefficients,
 \begin{align}
    H_5^{\rm Phys}(x)= &
    ( x^4+2 x^5+2 x^6+4 x^7+3 x^8) / D_5(x)  \nonumber\\
    &+ \frac{(x^9+x^{10})}{(1-x^2)(1-x^4)(1-x^5)(1-x^6)} \,,
     \label{5pt Hilbert series physical}
\end{align}
but this decomposition is not unique. Modulo the primary invariants of the denominators, this series consists out of 14 different hook structures. However, one of these can be written in terms of a secondary invariant, implying that there are 13 independent hook structures that can be used as 5-point seed interactions. 

\section{BCJ bootstrap}

\noindent 
From {\bf six-point} on, a systematic classification of scalar numerators becomes more complicated. The 6-point Molien series for invariants is 
\begin{align}
    H_6^{\rm Inv}(x) =&  (1 + 2 x^5 + 5 x^6 + 7 x^7 + 9 x^8 + 11 x^9 + 13 x^{10} \nonumber \\
    & + 14 x^{11}+ 21 x^{12} + 24 x^{13} + 28 x^{14} + 32 x^{15} \nonumber \\
    & + 26 x^{16} + 22 x^{17} +  13 x^{18} + 7 x^{19} + 3 x^{20} \nonumber \\
    & + x^{21} + x^{22})/ D_6(x) \,,
     \label{6pt Hilbert series invariant}
 \end{align}
 in terms of the following denominator containing the primary invariants:
  \begin{align}
    D_6(x)=& (1 - x^2) (1 - x^3)^2 (1 - x^4)^3 (1 - x^5)^2 (1 - x^6) \,. 
\end{align}
Thus there are nine primary invariants and 239 secondary ones. The Molien series of the BCJ irreps and gauge parameters can be establised in a similar manner. However, deriving the final number of independent structures requires the explicit forms of the primary and secondary invariants is problematic due to complicated relations (with syzygies of syzygies~\cite{benson_1993,Hilbert1890berDT}). We will not attempt such a general classification to all orders, and only list the number of physical and gauge parameters at lowest orders in table~\ref{table:6-point data}.

Moreover, we will focus on the subset of 6pt interactions that follow from 4pt seeds; in other words, we will require that they are BCJ bootstrappable from 4-point seed interactions, similar to \cite{Brown:2023srz}. This implies in particular that at singular channels such as $s_{abc}\rightarrow 0$, the amplitude should factorize into two 4-point amplitudes. In turn, this implies that a scalar numerator of ${\cal O}(s^p)$ factorizes as 
\begin{align}
    \lim_{s_{abc}\rightarrow 0}N_6^{(p)} = \sum_q c_q N_4^{(p-q)}(abcx) N_4^{(p+q)}(xdef)  \,,
\end{align}
where $x$ denotes the internal leg. As the lowest 4-point scalar numerator is linear in Mandelstam, the first $N_6$ that can be bootstrapped is quadratic. Beyond that, we find up to ${\cal O}(s^4)$:
\begin{align}
    \lim_{s_{abc}\rightarrow 0}N_6^{(2)}&=(s_{ac}-s_{bc}) (s_{de}-s_{df}) \,, \notag \\
    \lim_{s_{abc}\rightarrow 0}N_6^{(3)}&=(s_{ac}-s_{bc})  s_{ab} (s_{de}-s_{df})\nonumber\\
    &\hspace{7mm}+(s_{ac}-s_{bc})  (s_{de}-s_{df})s_{ef} \,, \notag \\
    \lim_{s_{abc}\rightarrow 0}N_6^{(4)}&=(s_{ac}-s_{bc})s_{ab}  (s_{de}-s_{df})s_{ef} \,.
\end{align}
Imposing this BCJ bootstrap fixes $N_6^{(2)}$ uniquely to
 \begin{align}
     N_6^{(2)} =& \left(s_{ac}-s_{bc}\right) \left(s_{de}-s_{df}\right) + \nonumber\\
     &+\tfrac{1}{2} s_{abc} \left(s_{ae}-s_{af}-s_{be}+s_{bf}\right) \,.
     \end{align}
while in other cases it still leaves some free parameters which can be seen as contact interactions that are separately BCJ compatible. We provide an overview of these numbers in table~\ref{table:6-point data}.

\begin{table}
\centering
\begin{tabular}{||c | |c | c | c| | c| | c ||} 
 \hline
$p$ & ${\rm BCJ}$ & ${\rm Phys}$ & ${\rm Gauge}$ & ${\rm Inv}$ & Bootstrap \\ [0.5ex] 
 \hline\hline
 1 &  1 &  1 & 0 & 0 & 0\\
 \hline
 2 &  3 &  3  & 0 & 1 & 1\\
 \hline
 3 &  9 &  8 & 1 &  2 & 2 \\
 \hline
 4 &  23 & 18 &  5 &  4  &  3\\
 \hline
 5 &  54 & 38 & 16 & 6 &  8\\
 \hline
 6 &  121 & 79 & 42 &  13 &  24\\
 \hline
 7 &  246 & 151 & 95 & 19 & 53 \\
 \hline
\end{tabular}
    \caption{\it The number of 6-point BCJ-compatible scalar numerators (split into physical and gauge parameters) and invariants at ${\cal O}(s^p)$. The last column lists the number of scalar numerators that are compatible with the BCJ bootstrap.}
\label{table:6-point data}
\end{table}

\section{Single copy: the gauged NLSM}
\label{sec:gauged NLSM}

\noindent
The systematic classification of seed scalar numerators at 4-point and the corresponding bootstrapped ones at 6- and higher-point allows for the construction of novel theories that feature interactions with different soft limits. As a first illustration, we will propose a single copy theory for an adjoint Goldstone scalar field, with amplitudes generated by the product of a colour factor with a linear combination of the different elementary solutions involving Mandelstam. Moreover, we will use the requirement of $\sigma_{\rm min} =0$ as a guiding principle. This will result in a gauged version of the chiral NLSM with symmetry breaking $G \times G \rightarrow G$, with additional interactions due to gluon exchange.

At four-point this theory is generated by $C_4 \times (N_4^{(1)} + N_4^{(2)})$, and therefore has two different contributions to the amplitudes. The corresponding Lagrangian is 
 \begin{align}
     \mathcal{L}_4 = - \tfrac12 (D \phi)^2 + \tfrac16 f^2 \phi^2 (D \phi)^2 - \tfrac14 F^2 \,,
 \end{align}
up to this order. 

Moving to six-point, we consider the schematic form $C_6 \times (N_6^{(2)} + N_6^{(3)} + N_6^{(4)})$. While the quadratic scalar numerator is unique, that is not the case for the cubic and quartic ones. To unambiguously determine the theory, we impose the soft limit $\sigma_{\rm min} = 0$ at cubic and $\sigma_{\rm max} = 1$ at quartic order; as promised in the introduction, this theory has different soft degrees with $\sigma_{\rm max} = \sigma_{\rm min} +1$. We have to extend the Lagrangian of the gauged NLSM with terms of the following form\footnote{A similar gauged NLSM was also considered in \cite{Carrasco:2022sck}, whose analysis also includes gluonic external states and concluded that the operator $D^2 F^4$ is needed. This theory can be thought as the dimensional reduction~\cite{Cheung:2017ems,Cheung:2017yef} of the ${\rm YM}+(DF)^2$ theory~\cite{Johansson:2017srf,Johansson:2018ues}. Note that our gauged NLSM is different since we only consider pion scattering instead. As a result, our theory only has a finite number of numerators at 4-point and is fixed by the required soft behaviours $\sigma_{\rm min/max}$ (in contrast to \cite{Carrasco:2022sck}). For further details, see \ref{sec:appendix}.}:
 \begin{align}
    \mathcal{L}_6 = \mathcal{L}_4 + \tfrac{1}{45} f^4 \phi^4 (D \phi)^2 - 2f^2 F^3 + \tfrac{1}{6}f^2\phi^2 F^2 \,,
 \end{align}
to generate these amplitudes correctly.

Moving to higher multiplicities, we conjecture that this pattern continues. For instance, at 8-point, one can take the amplitude generated by BCJ numerators of the form $C_8 \times (N_8^{(3)} + .. + N_8^{(6)})$. The corresponding Lagrangian will include all terms above plus $\phi^6 (D\phi)^2$ and possibly $F^4$ and $F^2\phi^4$. Note that all terms are gauge covariant, and will thus result in $\sigma_{\rm min} =0$ amplitudes. Moreover, the purely scalar two-derivative part reduces to the NLSM with $\sigma_{\rm max} = 1$. Thus one should think of this theory as the NLSM with subleading terms included. These are dictated by a combination of non-linear symmetries (e.g.~the structure of two-derivative terms), gauge invariance (i.e.~the covariant derivatives) and BCJ consistency (e.g.~the $F^3$ and $F^2 \phi^2$)\footnote{Note that the NLSM can also be fixed by imposing BCJ consistency and the exact pole structure, see \cite{Carrasco:2019qwr}.}.
 
\section{Double copy: DBI-Lovelock}
\label{sec:extended Goldstone}

\noindent
As a second example we propose a double copy theory that involves a single scalar field and that is fully determined by two different non-linear symmetries, with BCJ compatibility arising as a result.  This theory turns out to be the double copy of a gauged and an ungauged NSLM.

We again start from the full classification at 4-point. The scalar numerators $N_4^{(1)} \times N_4^{(2)}$ yields DBI with the quartic operator $(\partial \phi)^4$. In order to retain the $\sigma_{\rm min} = 2$ generalised Adler zero, one must add $(\partial \phi)^{2n}$ with specific coefficients at every order \cite{Cheung:2014dqa}. The full DBI theory is then given by the measure of 
 \begin{align}
    g_{\mu\nu} = \eta_{\mu\nu} + \partial_\mu \phi \partial_\nu \phi \,,
    \label{special metric}
 \end{align}
which can be seen as a brane-induced metric and is covariant under \cite{deRham:2010eu}
\begin{align}
  \delta \phi = c_{\mu}x^{\mu} + c_{\mu}\phi \partial^{\mu} \phi \,,
\end{align}
that generate a non-linear realisation of 5D Poincare symmetries.

The BCJ product $N_4^{(2)} \times N_4^{(2)}$, instead, yields the special Galileon (SG) theory, with operator $(\partial\phi)^2\left([\Pi]^2-[\Pi^2]\right),\;\Pi_{\mu\nu}=\partial_{\mu}\partial_{\nu}\phi$, $[\dots]={\rm Tr}[\dots]$, and the non-linear symmetry \cite{Hinterbichler:2015pqa}
 \begin{align}
  \delta \phi = s_{\mu\nu} x^{\mu}x^{\nu} + s_{\mu\nu} \partial^\mu \phi \partial^\nu \phi \,,
  \end{align}
resulting in the soft degree $\sigma_{\max} = 3$.

Can these be combined into a single, extended Goldstone theory that is BCJ-compatible based on the product $N_4^{(2)} \times (N_4^{(1)} + N_4^{(2)})$? We will provide evidence that the answer to this question is affirmative. The defining property, similar to DBI and SG, will be the soft limit: all interactions are required to have at least $\sigma_{\rm min} =2$. This is naturally satisfied when taking curvature invariants of the metric \eqref{special metric}; a general EFT would therefore be
 \begin{align}
  \mathcal{L} = \sqrt{-g} [ c_0 + c_1 R + c_2 R^2 + \ldots ] \,.
 \end{align}
However, this does not display the $\sigma_{\rm max} = 3$ scaling in any limit. In order to ensure that the highest-derivative terms have the SG scaling, one needs to restrict to the specific Lovelock invariants at every order; these have the special property of being degenerate and hence do not generate any corrections with more than two derivatives on a given field \cite{Lovelock:1971yv}. When evaluated at the induced metric \eqref{special metric}, the Lovelock invariants become total derivatives:
 \begin{align}
  R^n &\equiv \delta_{\alpha_1\beta_1\dots\alpha_n\beta_n}^{\mu_1\nu_1\dots\mu_n\nu_n}\prod_{i=1}^n R^{\alpha_r\beta_r}{}_{\mu_r\nu_r} = \partial_\mu j^{(n)\mu}\,.
 \end{align}
For illustration, the currents for $n=1,2$ are given by
\begin{align}
 j^{(1)\mu} = &\frac{2}{1+(\partial \phi)^2} \left(\phi^\mu[\Pi]  - \phi_\nu\Pi^{\nu\mu} \right) \,, \nonumber \\
 j^{(2)\mu} = & \frac{4}{\left(1+(\partial \phi)^2\right)^2} (2 \phi^\mu [\Pi^3]-3 \phi^\mu [\Pi^2][\Pi]  \nonumber \\
  + & \phi^\mu [\Pi]^3+3 \phi_\nu \Pi^{\nu\mu} [\Pi^2] -6 \phi_\kappa \Pi^\kappa_\lambda \Pi^\lambda_\nu \Pi^{\nu\mu}  \nonumber \\
  + & 6 \phi_\kappa \Pi^\kappa_\nu \Pi^{\nu\mu} [\Pi] -3 \phi_\nu \Pi^{\nu\mu} [\Pi]^2 ) \,, 
\end{align}
with $\phi_\mu=\partial_\mu \phi$. At lowest order in $\phi$, these operators become
 \begin{align}
 \sqrt{-g} R^n 
  &\simeq (\partial\phi)^2\delta_{\mu_1\dots\mu_n}^{\nu_1\dots\nu_n}\prod_{r=1}^n (\partial^{\mu_r} \partial_{\nu_r} \phi) \,,
 \end{align}
up to an overal constant.
Note that these are exactly the SG invariants; indeed, both the Lovelock and the SG invariants trivialise in sufficiently low dimensions $D$.

We can therefore tune the Lovelock coefficients to have a theory that propagates a single scalar field, has $\sigma_{\rm min}=2$ for all amplitudes (ensuring that it is DBI plus higher-derivative corrections) and moreover the higher-derivative amplitude has $\sigma_{\rm max} = 3$ (such that it asymptotes to the SG). This {\it DBI-Lovelock} theory is thus defined to all orders (and in all dimensions) uniquely by its non-linear symmetries and associated soft degrees.

We have checked that the amplitudes resulting from the above theory can be written in terms of a linear combination of scalar numerators, at least up to this order. Moreover, the specific linear combination of quadratic, cubic and quartic 6-point terms that are needed for the gauged NLSM and the DBI-Lovelock theory are identical.

\section{Conclusion}

This Letter opens up the tantalizing possibility that various Goldstone theories with hybrid soft degrees $\sigma$ adhere to the double copy paradigm. Moreover, these can be systematically classified using representation theory of $S_n$: at 4- and 5-point, respectively, there are two and 13 scalar numerators in terms of Mandelstam variables. 

Focussing on the former, we have identified a new sector of scalar field theories that can be BCJ-bootstrapped, over and beyond the analysis of \cite{Brown:2023srz}: the inclusion of the linear 4-point seed interaction introduces lower- (instead of higher-) derivative corrections to the NLSM. These allow for the construction of a gauged NLSM. BCJ compatibility then requires specific $F^3$ and $\phi^2 F^2$ terms. Moreover, the two 4-point seeds can be double copied into an extension of the special Galileon theory with lower-derivative corrections, which can be phrased in terms of Lovelock invariants of the DBI metric.

We have demonstrated the bootstrap construction of our two example theories explicitly at 6-point\footnote{Naturally, it would be very interesting to verify that this behaviour persists at higher-point; we leave this for future work.}.  In both cases, the theories are strongly constrained by two soft degrees with $\sigma_{\rm max} = \sigma_{\rm min} + 1$: this plays a crucial role in uniquely determining the scalar numerators. Our theories therefore appear to be closely related to the so-called extended DBI theory\footnote{The same reference also includes the combination of gluon exchange with cubic interactions of bi-adjoint scalars; this theory can be understood along similar lines as the extended DBI theory.} \cite{Cachazo:2014xea}, which involves the NLSM and DBI in specific limits. Indeed this theory can also be phrased in terms of BCJ numerators \cite{Low:2020ubn}, and only has a finite number of operators contributing to $n$-point amplitudes. Moreover, the on-shell constructability of this kind of theories follows from the graded soft theorem proposed in \cite{Kampf:2021tbk}. 

Based on the current results, it thus appears there is a fundamental difference between the higher-derivative corrections of \cite{Carrasco:2022lbm, Chen:2023dcx,Brown:2023srz} on the one hand, and hybrid Goldstone theories such as extended DBI, the gauged NLSM and DBI-Lovelock on the other hand: the latter category does not require an infinite set of higher-derivative corrections at given multiplicity. We expect that this difference arises due to the special nature of the highest-derivative terms in these theories: these have a softer degree $\sigma_{\rm max}$, are related by the non-linear symmetry and are separately BCJ compatible. In contrast, the leading terms of the higher-derivative corrections $F^3$ and $F^4$ in \cite{Carrasco:2022lbm} (at vanishing gauge coupling, i.e.~of the form $(\partial A)^n$) do not have a non-linear symmetry (beyond Abelian gauge symmetry) and are not separately BCJ compatible. A similar discussion applies to the scalar BCJ bootstrap \cite{Brown:2023srz} and the KLT kernel bootstrap \cite{Chen:2023dcx}, whose 4-point seed structures do not include the linear exchange interaction \eqref{4pt_win_1}.  

Finally, the general analysis of this paper suggest a number of novel theories beyond the two examples that we outlined. At four-point, one can instead take the product $N_4^{(1)} \times (N_4^{(1)} + N_4^{(2)})$ leading to a combination of DBI with gravitational interactions; moreover, this can be extended to have an $SO(N)$ flavour along the lines of \cite{deNeeling:2022tsu}. More generally, one can consider the case where both BCJ numerators have multiple terms of different order in Mandelstam. At first sight one might expect this to lead to a theory with three soft degree sectors; it would be interesting to investigate how this relates to the graded soft theorem of \cite{Kampf:2021tbk} that only allows for $\sigma_{\min}$ and $\sigma_{\rm max}$ to differ by one. We leave this question to future research.

\section*{Acknowledgements}

\noindent
The authors would like to thank Tom\'{a}\v{s} Brauner, John Joseph Carrasco, Dijs de Neeling and Karol Kampf for very valuable discussions.

\bibliography{BCJNLSM.bib}

\appendix
\section{Supplementary Material}

\label{sec:appendix}

In this Supplementary Material, we provide additional details to support and clarify two points discussed in the main text. Firstly, we elaborate on the counting and construction of gauge BCJ numerators. 
{Secondly}, we compare our gauged NLSM to the dimensional reduction of the $F^2+(DF)^2$ theory, address the differences, and comment on the Cachazo-He-Yuan (CHY) formulation.

\subsection*{Representation of gauge numerators}

\noindent
As is mentioned in the main text, for a BCJ numerator to correspond to a gauge transformation (and hence generate a vanishing contribution to the amplitude) it has to be proportional to a Mandelstam variable, e.g.~at four- and five-point 
 \begin{align}
     N_{abcd} = s_{ab}G_{abcd} \,, \quad  N_{abcde} = s_{ab}G_{abcde} -s_{de}G_{edcba}\,, \label{decomp}
 \end{align}
times an additional tensor $G_{abc\dots}$ that lives in specific representations of $S_n$ that are listed in table II of the main text. 

The above is a necessary and sufficient condition for cancellation in the amplitude. To see this, observe that the amplitude in the double copy formulation is a sum over trivalent diagrams with different pole structures. In order for the amplitude to vanish, different channels have to collaborate. In fact, 
the collaboration occurs when two diagrams differ in at most one propagator, and the above form \eqref{decomp} then allows for cancellations.

This statement can be illustrated at low multiplicities. The four-point partial amplitude from the DC is
\begin{align}
\frac{N_{adbc}}{s_{ad}}-\frac{N_{abcd}}{s_{ab}} \,.
\end{align}
As there is no common propagator between the two channels, the decomposition \eqref{decomp} is a necessary condition. The vanishing of the amplitude is then achieved by $G_{abcd}=G_{adbc}$.  This linear equation, combined with the Jacobi identities for $N$, restrict $G$ to live in the $[1,1,1,1]$ irrep. 
At five-point, the partial amplitude reads
\begin{align}
    A_5(1,2,3,4,5) = & \frac{N_{abcde}}{s_{ab} s_{de}}+\frac{N_{bcdea}}{s_{ae} s_{bc}}+\frac{N_{cdeab}}{s_{ab} s_{cd}}\nonumber\\
    &+\frac{N_{deabc}}{s_{bc} s_{de}}+\frac{N_{eabcd}}{s_{ae} s_{cd}} \,.
    \label{eq: PA5}
\end{align}
These can be paired based on their common poles. For instance, the first term yields under \eqref{decomp}
\begin{align}
\frac{N_{abcde}}{s_{ab} s_{de }}=\frac{G_{abcde}}{s_{ab}}-\frac{G_{edcba}}{ s_{de }} \,,
\label{eq: 5pt gauge factor}
\end{align}
and there are two terms in \eqref{eq: PA5} that have a common propagator, i.e.
\begin{align}
    \frac{G_{cdeab}}{s_{ab}}\subset \frac{N_{cdeab}}{s_{ab}s_{cd}},\quad \frac{-G_{cbaed}}{s_{de}}\subset\frac{N_{deabc}}{s_{bc}s_{de}} \,.
\end{align}
Vanishing of the amplitude therefore imposes linear relations between the different components of $G$, which have an 
$S_n$-group-theoretic interpretation. Together with Jacobi identities on the numerators $N$, they require $G$ to lie in specific irreps of $S_n$ as listed in Table II of the main text.

The number of such gauge parameters can then be obtained by the Molien series of the corresponding irrep(s); for instance, as given by (17) of the main text at five-point. As different tensors $G$ can result in the same numerators $N$ (the mapping \eqref{decomp} from $G$ to $N$ is not necessarily injective), this serves as an upper bound for the number of BCJ gauge numerators. The latter has the same denominator as the Molien series\footnote{This also follows from the fact that a polynomial ring, whose elements are in some irrep of $S_n$, should be generated by the product of a polynomial module and a polynomial ring. The generators of the module live in that specific irrep and the ring is a polynomial ring of invariants under $S_n$. {Such a decomposition of a polynomial ring is assured by the Cohen-Macauley property of an invariant polynomial ring under $S_n$ or a subgroup of $S_n$ \cite{derksen2015computational}.}} but can have smaller coefficients in its numerator. At five-point, we have explicitly constructed BCJ gauge numerators up to ${\cal O}(s^{13})$ and in this way determined (18).

\subsection*{Details of the gauged NLSM}

\noindent
Finally, we would like to discuss specific details of our gauged NLSM (26), and in particular compare it to the dimensional reduction  of the $(DF)^2+{\rm YM}$ theory \cite{Carrasco:2022sck}. The two theories have scalar amplitudes that are similar but differ in crucial aspects, including
\begin{itemize}
\item
The gauged NLSM of \cite{Carrasco:2022sck} can be derived as a dimensional reduction of $(DF)^2+{\rm YM}$ theory \cite{Johansson:2017srf,Johansson:2018ues} over a circle. This results in an infinite tower of scalar amplitudes with increasing number of derivatives at a given multiplicity. Instead, the gauged NLSM that we constructed truncates at a given derivative order; for instance, the 4-point amplitudes only has two orders, the gluon-exchange one and the NLSM one. In this sense, the two gauged NLSM are already different at 4-point. 
\item The gauged NLSM constructed in the main text is fixed by the soft degree. At the highest power of momentum, the amplitudes are designed to have the Adler's zero ($\sigma=1$). Instead, in \cite{Carrasco:2022sck}, there is a new contribution at the same order as the NLSM. This new contribution (Eq.~(5.3) in \cite{Carrasco:2022sck}) reduces the soft degree to $\sigma=0$. As a result, the gauged NLSM in \cite{Carrasco:2022sck} is not upper-bounded by Adler's zero, which might be related to the infinite tower of operators in that theory.
\end{itemize}
We therefore conclude that the gauged NLSM in our manuscript is different from the one in \cite{Carrasco:2022sck}. Of course, given the fact that the landscape of kinematic numerators of scalar theories is large (as illustrated in the main text by the counting of BCJ consistent numerators up to and including six-point), it is natural to expect that there exist multiple possibilities that pions can couple to gluons; in other words, there can be more than one gauged NLSM that is BCJ-compatible in its scalar sector, and it is valuable to explore these different possibilities.

We would also like to make a brief remark about the possible CHY formulation \cite{Cachazo:2013gna,Cachazo:2013hca,Cachazo:2013iea,Cachazo:2014nsa,Cachazo:2014xea} of the gauged NLSM(s). In the generalization of \cite{Cachazo:2016njl}, mixed amplitudes with external states belonging to different species can also be formulated in terms of CHY half-integrands, by combining the CHY building blocks of each species. However, in our view this does not straightforwardly carry over to the present situation, where the theory is about one type of particle with interaction terms of different momentum orders. In our gauged NLSM, the quadratic BCJ numerator at six-point corresponds to the dimensional reduction of YM and the cubic one follows from the dimensional reduction of the ${\cal O}(\alpha')$ contribution of $F^2+(DF)^2$. Instead, the quartic numerator is given by the NLSM. It would be interesting to check whether this pattern of combining dimensionally reduced $F^2+(DF)^2$ with the NLSM continues at higher multiplicities. If true, the CHY half-integrand of  $F^2+(DF)^2$ \cite{He:2018pol} can be used to produce the half-integrand of our gauged NLSM at all but the highest momentum order.

\end{document}